# Glass transition with decreasing correlation length during cooling of $Fe_{50}Co_{50}$ superlattice and strong liquids.

Shuai Wei[1], Isabella Gallino[1], Ralf Busch[1] & C. Austen Angell[2]

[1]Materials Science and Engineering Department, Saarland University, Universität Campus, 66123 Saarbrücken, Germany

[2]Department of Chemistry and Biochemistry, Arizona State University, Tempe, AZ 85287, USA

**The glass transition GT is usually thought of as a structural arrest that occurs during the cooling of a liquid, or sometimes a plastic crystal, trapping a metastable state of the system before it can recrystallize to stabler forms[1]. This phenomenon occurs in liquids of all classes, most recently in bulk metallic glassformers[2]. Much theoretical interest has been generated by the dynamical heterogeneity observed in cooling of fragile liquids[3,4], and many have suggested that the slow-down is caused by a related increasing correlation length[5-7]. Here we report both kinetics and thermodynamics of arrest in a system that disorders while in its ground state, exhibits a large $\Delta C_p$ on arrest ($\Delta C_p = C_{p,mobile} - C_{p,arrested}$), yet clearly is characterized by a correlation length that is decreasing as GT is approached from above. We show that GT kinetics in our system, the disordering superlattice $Fe_{50}Co_{50}$, satisfy the kinetic criterion for ideally "strong" glassformers[8], and since $\Delta C_p$ behavior through $T_g$ also correlates[8], we propose that very strong liquids and very fragile liquids are distinguished by existence on opposite flanks of an underlying order-disorder transition. For special conditions, this transition can become a liquid-liquid critical point but, as known in model systems of the Jagla type[9], it can also be either continuous or have weak first order character.**

Part of the difficulty of understanding the nature of viscous liquids and their structural arrest stems from the widely differing characteristics of the glass transition. While all glasses transitions are trivially associated with the crossing of experimental and system internal time scales, the physical signatures of this "ergodicity-breaking" event range from dramatic to almost invisible. For instance, in some cases the heat capacity drops to a mere 30% of its mobile state value while in others the drop is almost imperceptible. Among the latter cases are the archetypal glassformer, $SiO_2$ (dry), and its weak-field ionic cousin, $BeF_2$.

Silica is the extreme member of a general "strong/fragile" pattern of viscous slowdown in glassformers, a pattern which is also found for their thermodynamics [10]. For the fragile liquids, the rate at which the relaxation time increases on cooling is much greater than expected from the Arrhenius law, and the heat capacity jump at $T_g$ is large even when scaled by the excess entropy at $T_g$[10] (or by the more readily available entropy of fusion[11]).

Because of the unusual and extreme features of fragile liquids, these have received much more attention, from theoreticians and experimentalists alike, than have strong liquids and much is known about their behavior. A clear pattern of increasing heterogeneity with decreasing temperature has been observed, both in experiment[4] (particularly near $T_g$) and in computer simulation[5] (necessarily far above the experimental $T_g$ for the same sort of system). As one approach to understanding the origin of the slow-down, many studies have been made to detect a correlation length that increases with decreasing temperature[5-7, 12, 13]. Both static and dynamic length scales have been under discussion and length scales have been deduced from observations of both spontaneous fluctuations (anomalous light scattering[14]) and field-induced fluctuations which are considered easier to measure[7]. Berthier and coauthors in particular have promoted, via an inequality (their Eq. (5)) linking a four point correlator to a measureable susceptibility)[7]) the assumption that, for molecular liquids, "dynamic heterogeneity is strongly correlated with enthalpy fluctuations" hence to the excess heat capacity of liquid over crystal. On the other hand, the overarching importance of static correlations has recently been stressed by Tanaka and coworkers[15] who demonstrate critical-like behavior in colloids at the fragile liquid extreme[16], though the correlations of importance are not in the density.

Less attention has been given to these questions for liquids at the strong extreme of behavior. Some motivation for the present work has been provided by the realization that, for strong liquids, the excess heat capacity varies with distance from $T_g$ oppositely to the case of fragile liquids. With assistance from computer simulations, it has recently been found that both $SiO_2$ [17] and, more clearly, $BeF_2$[18], have excess heat capacities that peak far **above** the melting point, while the very bad glassformer, water seems to be a strong liquid near its $T_g$ but a very fragile liquid at temperatures near and above its melting point[19, 20]. There should, therefore, be some interest in the possibility that a correlation length changes in the opposite direction in strong liquid cases, i.e., it increases with increasing temperature above $T_g$ and thus acts to oppose viscous slowdown.

To explore this possibility, and the much broader picture of the glassformer problem that it necessarily leads to, we have chosen an unusual, but we hope fruitful, approach. We study a non-liquid system, with a strong and easily-studied "glass" transition, in which it is known that there is a static correlation length that changes with temperature, and we show that this glass transition and the associated disordering kinetics have a great deal in common with the behavior of the liquids at the strong extreme of the "strong-fragile" liquid pattern. Thus we have taken up the study of an unusual binary metal superlattice system, $Co_{50}$-$Fe_{50}$,[8, 21], that was first studied in the first half of the last century when interest in crystalline order-disorder (λ) transitions was at its height. Atypically, the ordering of Co and Fe onto their respective simple cubic lattices during cooling of the alloy is arrested midway through the transition, with a large drop, some 30%, in heat capacity.

The effect of different cooling rates (and hence ergodicity-breaking temperatures) on this glass transition was studied[21] in 1943, and it is qualitatively the same as observed for ionic and molecular glassformers by Moynihan and coworkers[22], and particularly by Yue and coworkers[23]. In the present work we study this phenomenology using different thermochemical protocols to extract the relaxation kinetics of the disordering process. In separate experiments we use an anneal-and-scan method to obtain the form of the heat capacity in the previously hidden zone below the lowest scanning $T_g$, and show thereby that the equilibrium heat capacity follows the previously known "lambda" form, as indeed was to be expected. This confirms that glass transitions can occur in systems in which the heat capacity is of theoretically understood form. More important is that we can argue that the kinetics of ordering in systems of this type are similar to the kinetics of ordering in "strong" liquids so we can then argue that ordering in "strong" liquids, while obviously associated with increasing relaxation time, is associated with a *decreasing* correlation length. The reason for this expectation is obviously that lambda transitions, like other critical phenomena, have static correlation lengths that are diverging at the critical point. It is particularly relevant that the relation between heat capacities across $T_g$ and near a subsequent liquid-liquid critical point have recently been established for a model system[9] (see below).

The technique used in our study is simple differential scanning calorimetry DSC, as described in "Methods". We have studied the relaxation kinetics in different ways, using scans through $T_g$ in which the heating rates are the same as the original cooling rate, and both rates change from case to case, (the original Moynihan method), and have also applied the fictive temperature method of Wang et al[24], which uses a constant heating rate upscan after cooling at different rates. The latter avoids any need for $T_g$ determination constructions (maximum slope, etc) that might be susceptible to instrument lag, and we have found it the most reliable (see Fig. 2 caption). The definition of the fictive temperature from the heating scan is shown for the lambda transition case, in Figure 1, lower section. The Wang-Velikov method yields the fragility directly from the slope of the Arrhenius plot (Figure 2, insert). This shows that, within experimental uncertainty, the fragility is the ideal "strong" liquid value of 16, corresponding to simple Arrhenius kinetics, notwithstanding the accelerating configurational heat capacity seen in Figure 1.

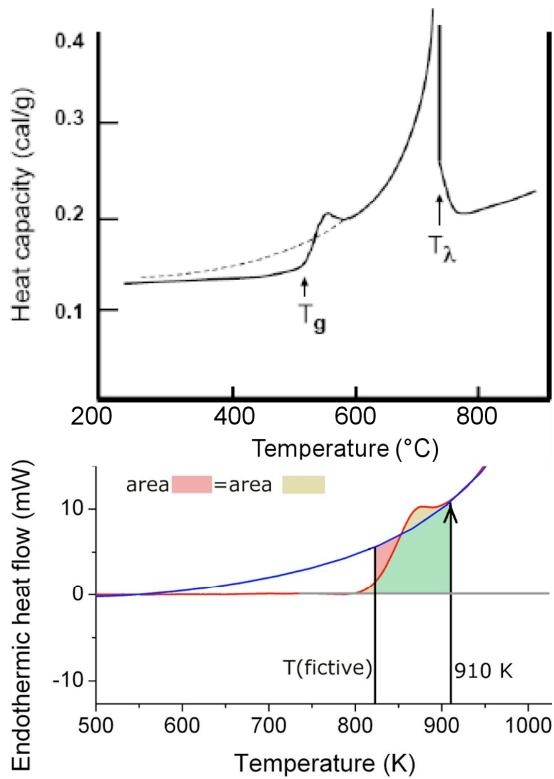

**Figure 1.** The glass transition in the $Fe_{50}Co_{50}$ order-disorder transition, showing lambda peak at 730ºC and ergodicity-remaking starting at about 500ºC, when effective heating rate is much greater than the 30K/day cooling rate. Note how the heat capacity jumps about 30% at the glass transition, (from about 10% above the classical value of 24.9 J/mol.K, to 1.4 $C_p$(classical)) (taken from ref. 18). The lower section shows the definition of the fictive temperature from Moynihan's equal area construction applied to $Fe_{50}Co_{50}$, using an upper temperature limit of 910K. The baseline is defined by the frozen sample heat capacity, for the case, heating rate = cooling rate = 20K/min.

In an parallel study, we have defined a relaxation time from the start, and end, values of the glass transition, as described by Busch et al.[25],

$$\tau = T_g(end) - T_g(start)/Q$$

where Q is the heating rate (in this method also the cooling rate) and found, again, simple Arrhenius behavior, now with pre-exponent $10^{-15}$ s. This is the ideally simple behavior seen in the "strong" limit of plastic crystals studied by Fujimori and Oguni[26] - one of which ($TlNO_2$) turns out to belong to the same lambda transition family as the present system (but one in which the ergodicity-breaking occurs only in the far tail of the transition, and is only detectable by precision adiabatic calorimetry).

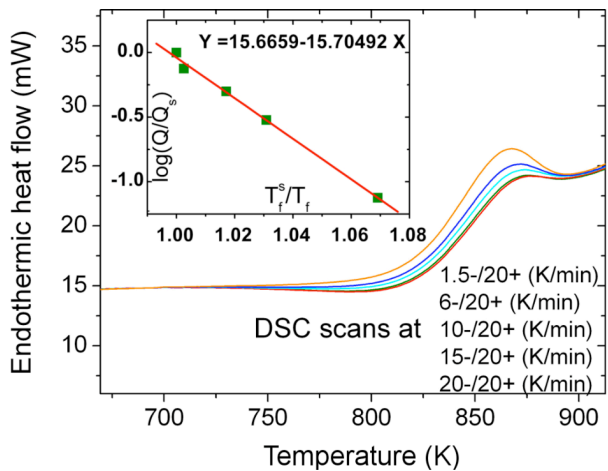

**Figure 2.** DSC upscans at fixed 20K/min rates, following cooling at different slower rates between 1.5 and 20K/min.. The Wang-Velikov method obtains the kinetics of the glass relaxation process by plotting the log of the ratio [sample cooling rate/ "standard" cooling rate (20K/min)] against the ratio of [standard fictive temperature ($T_f^s$)/sample fictive temperature], the latter being the value determined by the Fig. 1 construction for the chosen cooling rate. **Insert.** Wang-Velikov Arrhenius plot[24] in which both slope and intercept yield the "m fragility" (the slope of the Arrhenius plot of the relaxation time vs $T_g/T$) .

To establish that it is indeed the lower part of the normal lambda transition that is being chopped off by the glass transition of Figure 1, we have carried out annealing processes at lower temperatures followed by the enthalpy recovery scans, which reveal the enthalpy lost during the anneal (see SI). The attainment of equilibrium states was proven by demonstration of annealing time independence. The heat capacity derived from the difference between successive enthalpy recovery curve integrals, is shown in Figure 3. It is seen (1) that these form a natural continuation of the lambda form to lower temperatures, and (2) that this form can be well represented by the Kirkwood 2$^{nd}$ order approximation[27]. The Bragg-Williams approximation[27] predicts a heat capacity peak, but quantitatively is quite inadequate.

Numerical simulations widely available for the 2D Ising model, show that the correlations lengths for density and enthalpy fluctuations, that determine heat capacity, relaxation times, etc. are all increasing as $T_\lambda$ is approached both from above and from below. The latter is to be emphasized: the correlation length increases as the temperature moves away from the ergodicity-breaking temperature, $T_g$. This is exactly the opposite trend from that assigned to both static and dynamic correlation lengths of fragile glassformers as they approach their $T_g$'s from above. It is, on the other hand, the trend that must apply near $T_g$ to the recently-studied attractive Jagla model[9]. This model has has a liquid-liquid critical point in the stable liquid domain[28] so can be investigated at leisure without fear of crystallization.

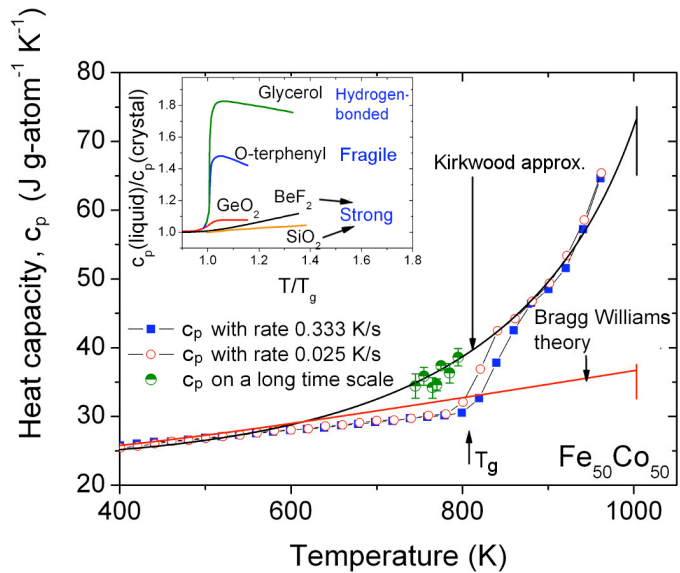

**Figure 3.** Comparison of equilibrium heat capacities, derived from the scans of enthalpy recovery (shown in the SI), with data obtained from fixed scan rate runs and also the theoretical functions from the Bragg-Williams model and the Kirkwood approximation. Clearly if the energy barrier opposing the elementary lattice site exchange were smaller, the glass transition in the alloy would occur at lower temperature and would quickly become difficult to detect, as in the case of most lambda transitions and also of dry vitreous silica, and vitreous water and. **Insert:** $C_p$ behavior of some liquid-based glasses through $T_g$.

The high density liquid phase in this latter system can be studied over wide temperature ranges as the temperature rises above the glass temperature, at pressures that are arbitrarily close to the critical pressure[9]. A heat capacity spike that has slow kinetics, and undoubtedly also, large static correlation lengths, has been described[9] and has been observed to damp out as the pressure departs increasingly from the critical value. Clearly, this correlation length is increasing as the spike is approached from below, just as surely as it is increasing as the spike is approached from above. The peak in heat capacity is reached as the system crosses the "Widom line"[29] that extends above the critical point as a higher-order continuation of the (first order transition) coexistence line that is terminated by the critical point (see the example for argon in Fig. 4 below). At $T > T_{(Widom)}$ i.e. above the $C_p$ peak, the liquid behaves in a very fragile fashion, partly because both correlation length effects and intrinsic kinetic effects are changing in the same direction. (The intrinsic kinetic factor is the energy barrier opposing the activated local rearrangements that are the elementary excitations in the hierarchical relaxation process[30]).

Below the Widom line crossing temperature, the static correlation length must decrease with decreasing temperature, like the case we are describing in the present study. We suggest that this effect opposes the intrinsic kinetic effect, leading to strong liquid behavior. The fragile-to-strong transition has been discussed in some detail for the case of bulk water[19, 31, 32], and in more detail for water in confined geometries[33-35] (where crystallization does not occur, but where water-water interactions related to those in bulk evidently persist). The increasing static correlation length that accompanies the diverging heat capacity as unconfined water supercools - long a matter of controversy[36, 37] - has recently been confirmed by Nilsson and coworkers, using x-ray scattering methods[38], now extended to -20°C[39].

Since the attractive Jagla model is indubitably a liquid, the following question naturally arises. Is the phenomenology we have just described at the core of the behavior of strong liquids like silica, $GeO_2$ and $BeF_2$? We have already referred to the peaks in heat capacity for $BeF_2$ and $SiO_2$ when high T data from simulations are included, and now give more detail. In particular, the heat capacity vs temperature calculated at different fixed frequencies, in the study of Binder and coworkers[17], shows an increasing "jump" as the frequency is increased (and the temperature of the jump moves higher). This is similar to the behavior observed in the present alloy system as the scan rate is increased. Furthermore, the temperature dependence of the enthalpy relaxation time, obtained from the temperature at which the imaginary part of the $C_p$ function exhibited its peak value, reportedly follows an Arrhenius law with activation energy 5.6 eV (132 kcal/mole)[17] in remarkable acccord with the activation energy for laboratory viscosity (134 kcal/mol[40]).

Finally, in the simplest of the $SiO_2$ models, it was observed[41] that, in a family of P vs density isotherms, the low temperatures members were beginning to inflect, i.e. the behavior was trending towards a liquid-liquid critical point - as in water and the more metal-like (more highly co-ordinated[9]) Jagla models. Certainly a fragile-to-strong transition, consistent with an off-critical Widom line crossing, has been identified in the kinetics of $SiO_2$ in the BKS model[42]. Certainly also, many metallic glassformers are currently revealing fragile-to-strong transitions[43, 44] (see SI) and corresponding thermodynamic signatures can be deduced.

To provide quantitative diagrammatic support for this scenario we present, in Figure 4, heat capacity data for the case of $BeF_2$ above its melting point[18], and compare it (and also $SiO_2$ [17]) with the data on the present system, and with data for confined water far below 0°C[20] - and finally with data for the off-critical heat capacity of an inert gas near its gas-liquid critical point[45]. The similarity of the 11.82 MPa argon data and that of the $BeF_2$ and $SiO_2$ high temperature liquids is not easily ignored. The implication of this comparison is clearly that the behavior of the classical inorganic network glasses is to be understood in terms of off-critical (or "narrowly avoided critical") point phenomenology, lattice-gas models for the critical point having much in common with models for lambda transitions in solids (see SI). Furthermore, and of obvious technological significance, the static correlation lengths for density (refractive index) fluctuations in $SiO_2$ glass must be a decreasing function of fictive temperature if we are correct.

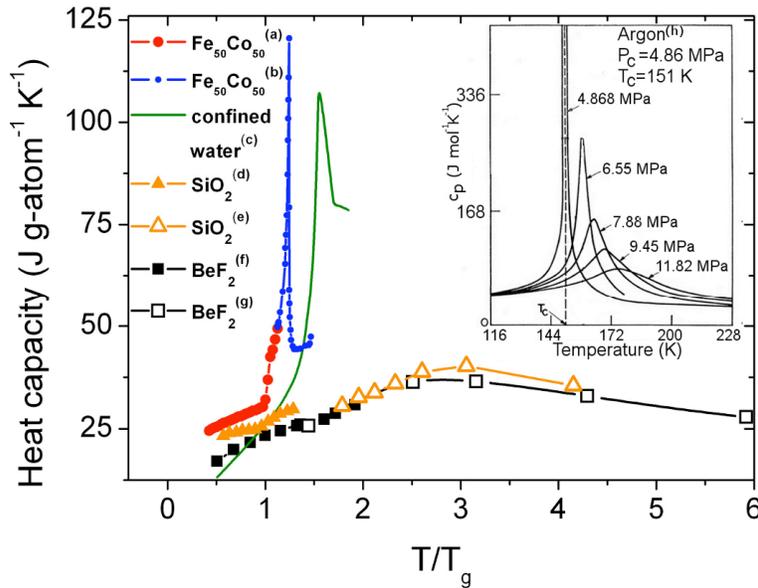

**Figure 4.** Heat capacities of $BeF_2$ and $SiO_2$ per g-atom above $T_g$ compared with that above $T_g$ for the CoFe alloy, and for confined water. The data sources a--g are fully documented in the Supplementary Information.
**Insert:** The same smearing to high temperature is seen for the off-critical pressure behavior of laboratory argon near its gas-liquid critical point (see SI).

The latter point can be checked. Although the correlation lengths are orders of magnitude smaller than the wavelength of laser light, the scattering of the latter will be affected through the Fourier components - and indeed, reduced light scattering in $SiO_2$ core glass fibres has been correlated with lower fictive temperatures[46]. It is likewise observed that, for confined water, the dynamic susceptibility $\chi_T(Q,t)$, determined by neutron scattering[35] and linked to the correlation length for dynamic heterogeneity[7, 35], decreases for temperatures below the heat capacity peak - though that for the corresponding 4-point correlator does not. The latter is evidently the case for $SiO_2$ also[47].

In summary, the study of the glass-like transition in the $Co_{50}Fe_{50}$ superlattice lambda transition provides an experimental parallel to the recent reports of glass transitions on the *low temperature* flank of the supercritical heat capacity spike in a model liquid system (Jagla model) that possesses a liquid-liquid critical point. Both provide examples of glass transitions with large heat capacity signatures, accompanied by correlation length decreases as $T_g$ is approached. There is a strong suggestion that this is related to "strong" liquid behavior in the classical network glassformer systems, which then implies that strong and fragile liquids exist on opposite flanks of an underlying order-disorder transition. In the very fragile liquid cases, according to theoretical fittings of excess $C_p$ and entropy data[30], this continuous process is interrupted by a first order transition to the low entropy state that occurs a little below $T_g$ (i.e. on long observation times), as discussed elsewhere[8, 30] (see SI). The character of the underlying transition is left unresolved in the current critical ordering analysis of ref [15].

**Methods**
The $Fe_{50}Co_{50}$ alloy was prepared by arc melting ultrasonically cleansed Fe and Co pieces, of purity of 99.97% and 99.95%, respectively, in a high-purity argon atmosphere. After remelting under argon and casting into waterchilled copper molds (5mm bore), the as-cast alloy rod was cut into 1mm discs, using a diamond saw, and sealed into high purity gold DSC pans.
Calorimetry was performed using a ***Perkin-Elmer Diamond Differential Scanning Calorimeter*** under constant flow (20 ml/min) of high-purity argon. For the fictive temperature determinations, cooling rates were chosen not to exceed the fixed 20K/min heating rate. Data points obtained at cooling rates greater than 20K/min in the fictive temperature measurements (and at scan rates greater than 20K/min in the relaxation time measurements) were considered unreliable, due to insufficient temperature equilibration, and have not been included in the assessment of the kinetics.

**Acknowlwedgements**
We appreciate support received from the Deutsche Forschungsgemeinschaft (DFG). CAA acknowledges helpful discussions with Mark Ediger and Juanzheng Yue.